\PassOptionsToPackage{numbers,sort&compress}{natbib}
\documentclass[final,5p,times,twocolumn]{elsarticle}
\journal{Physics Letters B}

\usepackage{preamble}
\usepackage{slashed}
\usepackage{braket}
\usepackage[nameinlink]{cleveref}

\def\sect#1{\hyperref[#1]{\Cref{#1}}}
\Crefname{section}{Sec.}{Sects.}
\Crefname{figure}{Fig.}{Figs.}
\Crefname{table}{Tab.}{Tabs.}

\begin{document}

\begin{frontmatter}
\title{Baryonic form factors of light pseudoscalar mesons}

\author{A.S.~Miramontes}
\ead{angel.s.miramontes@uv.es}
\author{J.M.~Morgado}
\ead{jose.m.morgado@uv.es}
\author{J.~Papavassiliou}
\ead{joannis.papavassiliou@uv.es}

\affiliation{
  organization={\mbox{Department of Theoretical Physics and IFIC, University of Valencia and CSIC}},
  city={Valencia},
  postcode={E-46100},
  country={Spain}
}

\begin{abstract}
Employing the Bethe-Salpeter formalism, we present a computation of the space-like baryonic form factor for the pion and kaon. In the exact isospin-symmetric limit this observable is forbidden by $G$-parity, so that any nonzero signal constitutes a direct probe of the quark mass difference $m_d - m_u$. The form factors are evaluated in the impulse approximation using fully dressed quark propagators, meson Bethe-Salpeter amplitudes, and a dressed baryon-current vertex
constrained by the vector Ward-Takahashi identity. The baryonic radius computed with this method 
for the pion is given by $\langle r_{\! B}^2\rangle_{\pi^+}^{1/2} = 0.043(2)$ fm, and is  consistent
with the available dispersive benchmarks. 
Our predictions for the kaons, namely 
$\langle r_{\!B}^2\rangle_{K^+}^{1/2} = 0.265(7)$ fm and 
$\langle r_{\!B}^2\rangle_{K^0}^{1/2} = 0.262(7)$ fm, 
indicate a larger spatial extent than in the pion case; these results 
have no dispersive counterparts, and are 
compatible with chiral QCD models. 

\end{abstract}

\end{frontmatter}

\section{Introduction}\label{sec:intro}

Electromagnetic form factors provide a direct window on the internal structure of hadrons, encoding the electric charge  distribution as a function of momentum transfer.
Less familiar, but conceptually analogous, are the form factors associated with other conserved vector charges of QCD, which probe complementary aspects of the hadron structure.
Among them, the form factor linked to baryon number 
is a particularly clean observable in the light-meson sector.
In particular, 
for a $q\bar q$ state, the total baryon number is zero; therefore, in the limit of exact isospin symmetry, 
the corresponding pion form factor is forced to 
vanish identically for all momenta.
Consequently, any nonzero signal is driven entirely by  
isospin breaking, induced by the quark mass difference
$m_d-m_u$, as well as by QED effects. Thus, the "baryonic" form factor (BFF)
serves as a sensitive indicator of how the quark and antiquark contributions fail to cancel each other inside a meson.

The first quantitative study of this observable was carried out in \cite{Sanchez-Puertas:2021eqj} for the charged pion; there, a constituent quark model estimate was complemented by a data-driven extraction from high-statistics $(e^+e^-)$ data collected by BaBar \cite{BaBar:2009wpw} and KLOE \cite{KLOE:2004lnj, KLOE:2008fmq, KLOE-2:2017fda, KLOE:2010qei}, analytically continued to the space-like regime via dispersive relations. That analysis yielded a small but definite baryonic radius, and provided a phenomenological determination of the form factor as a function of the momentum transfer. The kaon case was subsequently explored in \cite{Broniowski:2021awb}, employing chiral quark models.

In this work, we compute the BFFs of the $\pi^{+}$, $K^{+}$ and $K^{0}$ within the framework of Schwinger-Dyson (SDE) and Bethe-Salpeter (BSE) equations \cite{Salpeter:1951sz,PhysRev.84.350,Bethe1957,Nakanishi:1969ph,Jain:1993qh,Munczek:1994zz}, employing a flavour-dependent effective interaction.  The SDE-BSE methods have been extensively used in the investigation of hadron spectroscopy and structure \cite{Maris:1997tm, Alkofer:2002bp, Nicmorus:2008vb, Hilger:2015hka, Hilger:2014nma, Eichmann:2016yit,Mojica:2017tvh, Serna:2017nlr,Bednar:2018mtf,Wallbott:2019dng,Miramontes:2022mex,Liu:2023reo,Chen:2023zhh, Tandy:2023zio,Hoffer:2024fgm, Hoffer:2024alv,Miramontes:2025imd, Ferreira:2025wpu, Hagel:2025ngi, Eichmann:2025wgs, Huber:2025cbd, Alkofer:2026vux}. In particular, they have been successfully applied to the description of light-meson electromagnetic form factors \cite{Maris:1999nt,Maris:1999bh, Cloet:2008re,El-Bennich:2008dhc, Sanchis-Alepuz:2013iia, Weil:2017knt, Eichmann:2019tjk, Eichmann:2019bqf, Miramontes:2021xgn, Miramontes:2021exi, Raya:2015gva, Miramontes:2025ofw} and, more recently, have been applied to heavy-light systems \cite{Qin:2019oar, Xu:2024fun, Xu:2024vkn, Gao:2024gdj, Miramontes:2025vzb}.

The BFFs are constructed from the baryon-number current coupled to the underlying $q\bar q$ dynamics through a consistent set of ingredients: dressed quark propagators, meson Bethe-Salpeter (BS) amplitudes, and a dressed baryon-current vertex. 
This particular vertex is  
composed of twelve tensorial structures and the  
associated form factors, which are determined 
from the corresponding SDE and the constraints 
imposed by the vector Ward-Takahashi identity (WTI).
Thus, the vertex employed in our analysis 
encodes dynamical strength beyond the  classical $\gamma^\mu$ insertion.

The article is organized as follows. In \sect{sec:framework} we summarize the general framework employed for the calculation of the form factors. In \sect{sec:results} we include our main results, namely the BFFs and baryonic radii. Finally, in \sect{sec:outlook} we present our conclusions.

\begin{figure*}[t!]
\centerline{%
\includegraphics[width=0.8\textwidth]{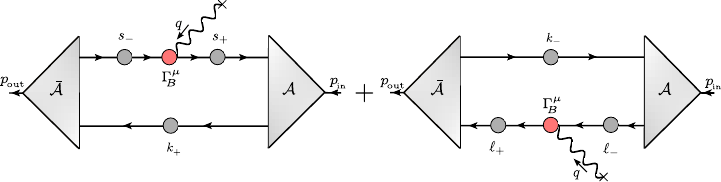}}
\caption{Diagrams of the baryon-number current in the impulse approximation that we employ. The wavy lines with a 
cross attached to them indicate the insertion of the baryon-number current, the orange circles represent the fully dressed current vertices, while the gray ones denote dressed quark propagators.}
\label{fig:impulse}     
\end{figure*}

\section{General framework}\label{sec:framework}

In this section, we summarize the formalism employed for the computation of the BFFs of the pion and kaons.

In order to fix the ideas, we recall that in QCD with exact isospin symmetry, $m_u=m_d$, and neglecting QED, the $G$-parity, $G=C\,e^{i\pi I_2}$, is an exact symmetry of the theory.
The baryon-number current, $j_{\! B}^\mu$, is odd under charge conjugation, whereas the pion is a $G$-parity eigenstate. As a consequence, the corresponding matrix element vanishes, and,
in that limit, $F_{\! B}^{\pi^+}(q^2) = 0$, 
for all momentum transfers~\cite{Khodjamirian:2020btr, SakuraiSakurai+2015}.
Once isospin symmetry is broken, this cancellation no longer holds, and a nonzero BFF is induced. In the following, we focus on the strong isospin-breaking contribution associated with $m_d-m_u$, leaving additional QED effects for a future, more 
detailed study.

From the practical point of view, the computation of a BFF for pseudoscalar mesons proceeds exactly as in the standard evaluation of electromagnetic form factors within the same SDE-BSE framework: one couples an external conserved vector current to the quark line through a fully dressed vertex, which obeys 
its own SDE. In particular, 
in the case of the electromagnetic current, one has 
\begin{equation}
  j_{\rm em}^\mu = \sum_f Q_f\,\bar q_f\gamma^\mu q_f\,,
\end{equation}
with $Q_u=+2/3$, $Q_d=-1/3$, etc.  The baryon-number current is instead flavour blind,
\begin{equation}
  j_{\! B}^\mu = \frac{1}{3}\sum_f \bar q_f\gamma^\mu q_f\,,
\end{equation}
so that the Dirac and momentum structure of the  vertex SDE are unchanged, and only the flavour weights associated with the external current are modified. 

The central ingredients for the calculation of the form factors are the dressed quark propagator, the meson BSE, and the SDE for the baryon-current vertex, which is formally identical to the quark-photon vertex equation computed in \cite{Miramontes:2025vzb}.

In the present work, we employ an explicitly isospin-breaking setup, and therefore solve separate quark gap equations for the $u$, $d$, and $s$ flavours,

\subsection{Baryonic form factor} \label{subsection:BFF}

The BFF of a pseudoscalar meson is defined from the matrix element of the baryon-number current between on-shell states. Denoting by $j_{\! B}^\mu(x)$ the conserved baryon current, and by $\ket{\textbf{s}(p)}$ the meson under consideration, with mass $M_\textbf{s}$, the hadronic matrix element may be written as
\begin{equation}
  \bra{\textbf{s}(p_{\mathrm{out}})} j_{\! B}^\mu(0) \ket{\textbf{s}(p_\mathrm{in})}
  = (p_{\mathrm{out}} + p_{\mathrm{in}})^\mu\, F_{\! B}^{\textbf{s}}(q^2)\,,
  \label{eq:JB-matrix-element}
\end{equation}
where $q = p_{\mathrm{out}} - p_{\mathrm{in}}$ is the momentum of the current, $p_{\mathrm{in}}$ and $p_{\mathrm{out}}$ denote the initial and final meson momentum, and $F_{\!B}^{\textbf{s}}(q^2)$ is the baryonic form factor
\cite{Sanchez-Puertas:2021eqj}. It is important to remark that in the isospin-symmetric limit, where $m_u =m_d$ and the electromagnetic effects are neglected, the total baryon number of the pion is exactly zero, and the corresponding form factor $F_{\!B}(q^2)$ vanishes. 
A non-zero baryonic form factor of the pion thus arises from isospin breaking effects, such as the quark mass difference  $m_d - m_u$.
Within the present framework, the matrix element of the baryonic current is evaluated in the impulse approximation, by coupling the current to the constituent quarks inside the meson via a fully dressed baryon-current, see \Cref{fig:impulse}. Specifically, 
\begin{equation}
  \begin{aligned}
J_{\!B}^\mu(p_{\text{av}},q) =
    & \int_k \, \bar{\mathcal{A}}_{f f'} (k_{\mathrm{out}}, p_{\mathrm{out}})\,
    S_{\!f}(\ell_+)\, \Gamma_{B,f}^\mu(k_+,q)\, S_{\!f}(\ell_-)\,
    \\ & \times \mathcal{A}_{f f'}(k_{\mathrm{in}}, p_{\mathrm{in}})\, S_{\!f'}(k_-) 
     + (f \leftrightarrow f')\,,
    \end{aligned}
  \label{eq:JB-impulse}
\end{equation}
where $\mathcal{A}$ is the meson BS amplitude, $\bar{\mathcal{A}}$ is the charge conjugated meson amplitude, and $S_{\!f}$ is the quark propagator of flavour $f$. 
In addition, $\Gamma_{\!B}^\mu$ denotes the
vertex that describes the coupling of the 
current to a quark-antiquark pair; for brevity we will refer to it as the "$B$-vertex". 

The integration measure and kinematic variables are defined as follows
\begin{equation}
    \int_k:= (2\pi)^{-4}\int d^4k \,,\\
\end{equation}
\begin{align}
k_\pm = k \pm \tfrac{1}{2}\,p_{\text{av}},
\qquad \ell_{\pm} = k_+ \pm \tfrac{q}{2}\, \qquad s_{\pm} = k_{-} \pm \tfrac{q}{2}\,
\label{eq:measure-kinematics}
\end{align}
Once the current has been computed, the baryonic form factor can be extracted from the contraction,
\begin{equation}
  F_{\!B}^{\textbf{s}}(q^2) = 
  \frac{J_{\! B}^\mu(p_{\text{av}},q)\, p_{\text{av}}^\mu}{2\, p_{\text{av}}^2}\,,
  \label{eq:FB}
\end{equation}
evaluated for on-shell external momenta, 
$p_{\mathrm{out}}^2=p_{\mathrm{in}}^2=-M_{\textbf{s}}^2$, with $p_{\text{av}}$ parametrized as \mbox{$p_{\text{av}} = \big(0,0,0,i \sqrt{M^2_{\textbf{s}} + q^2/4}\big)$}.
Finally, the baryonic mean squared radius of the meson is defined from the slope at the origin
\begin{equation}
  \langle r_{\! B}^2\rangle_{\textbf{s}}
  = -6\, \frac{d F_{\!B}^{\textbf{s}}(q^2)}{dq^2}\bigg|_{q^2=0}
  \label{eq:rB2}
\end{equation}
This quantity provides a quantitative measure of the baryonic spatial distribution induced inside the meson by isospin breaking.

\subsection{Dynamical equations} \label{sec:dynamic_eqs}

\begin{figure*}[t!]
\centerline{%
\includegraphics[width=0.7\textwidth]{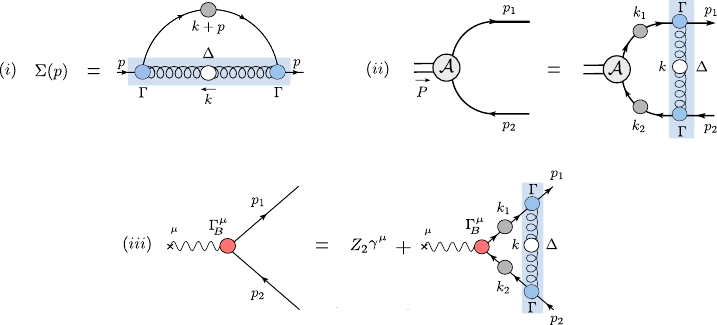}}
\caption{Diagrammatic representation of the main components: (i) the quark self-energy entering the gap equation; (ii) the meson Bethe-Salpeter amplitude for the quark-antiquark state; and (iii) the SDE for the baryon-current vertex. The blue band corresponds to the interaction kernel defined in Eq.~\eqref{eq:effective-gluon}.}
\label{fig:truncation}     
\end{figure*}

All building blocks entering the baryon-number current in \1eq{eq:JB-impulse} are obtained from their own dynamical equations, evaluated with the same flavour-dependent effective interaction used in Ref.~\cite{Gao:2024gdj}. In the present work we employ an explicitly isospin-breaking setup, and therefore solve separate quark gap equations for the $u$, $d$ and $s$ quarks, yielding distinct propagators $S_u$, $S_d$ and $S_s$. Consequently, the pion BS amplitude is computed as a {\it genuine} $u\bar d$ bound state (and analogously the kaons as $u\bar s$ and $d\bar s$), rather than in the isospin-symmetric basis.

The three dynamical equations 
relevant for the present  analysis 
are ({\it i}) the quark gap equation, 
({\it ii}) the meson BSE, and ({\it iii}) 
the SDE for the $B$-vertex, see \Cref{fig:truncation}.

\begin{figure*}[t!]
\centerline{%
\includegraphics[width=1.0\textwidth]{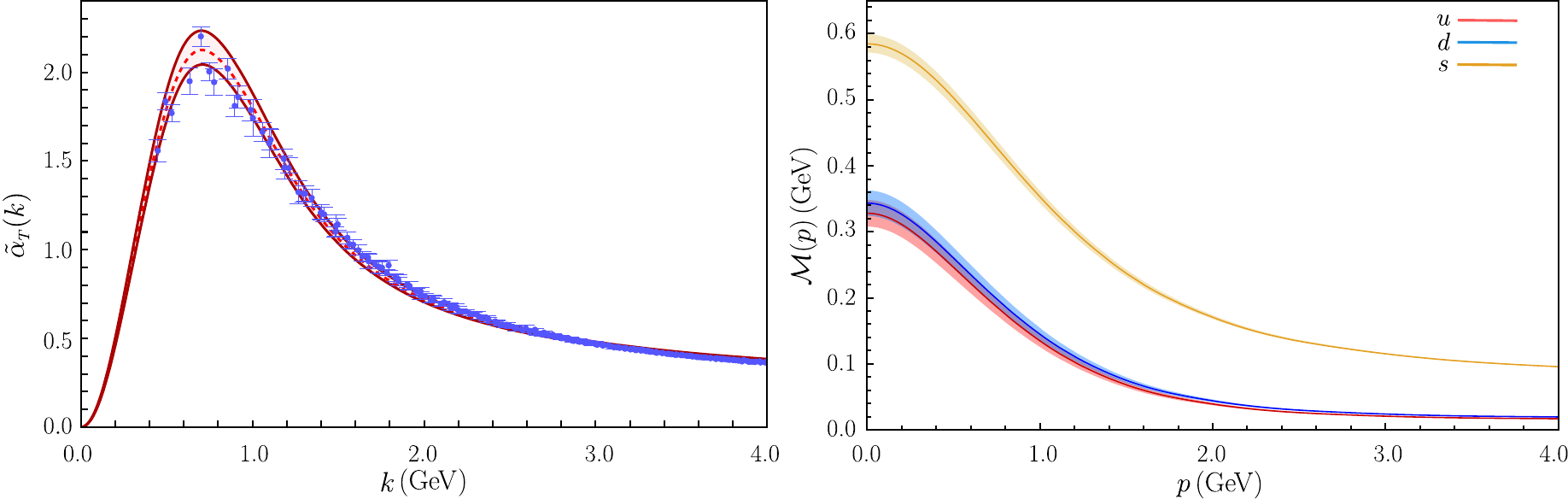}}
\caption{Left panel: The modified Taylor effective charge, ${\tilde\alpha}_{T}(q^2)$, The dashed curve represents the central parametrization from ~\cite{Gao:2024gdj}. The shaded band shows the variation obtained by shifting the fit parameters by one standard deviation ($\pm\sigma$) around their central values. Right panel: The mass function $\mathcal{M}_f(p^2)$ for the flavours $f=u,d,s$.}
\label{fig:aT_mass}     
\end{figure*}

Following \cite{Miramontes:2025vzb}, the 
key ingredient common to these three 
equations is
the flavour-dependent
effective interaction, $\mathcal{I}_{\! ff'}$$(k^2)$, defined as,
\begin{equation}
  \mathcal{I}_{\! ff'}(k^2) = \tilde{\alpha}_T(k^2)\, A_f(k^2)\, A_{f'}(k^2)\,,
  \label{eq:I_ff}
\end{equation}
where $\tilde{\alpha}_T(k^2)$ is the modified Taylor coupling introduced
in \cite{Gao:2024gdj}. The crucial 
difference between $\tilde{\alpha}_T(k^2)$ and the 
standard Taylor coupling $\alpha_T(k^2)$ \cite{vonSmekal:1997ohs,Boucaud:2008gn,vonSmekal:2009ae,Blossier:2012ef,Zafeiropoulos:2019flq} is that the former 
includes additional process-independent contributions, 
extracted from the two 
quark-gluon vertices attached to the 
gluon propagator (blue circles in \1eq{fig:truncation}). 
Convoluting these contributions with the lattice data for 
$\alpha_T(k^2)$ \cite{Zafeiropoulos:2019flq}, one 
finally obtains the $\tilde{\alpha}_T(k^2)$ shown in 
the left panel of \Cref{fig:aT_mass}.  There, 
the red dotted curve represents a convenient 
parametrization of $\tilde{\alpha}_T(k^2)$ introduced  
in ~\cite{Gao:2024gdj}, particularly suitable for the ensuing numerical 
treatment. On the same plot,
the red band corresponds to variations of $\tilde{\alpha}_T(k^2)$
obtained by shifting the fit parameters by one standard 
deviation. These variations, in turn, propagate into  
the quantities computed (quark propagator, pion BS amplitude, 
and BBFs), and can serve as error estimates of this approach. 

In the Landau gauge, this construction defines the effective gluon-exchange kernel
\begin{equation}
D_{\mu\nu}^{ff'}(k) = 4\pi\, \Delta_{\mu\nu}^0(k)\, \mathcal{I}_{ff'}(k^2)\,,
\qquad
\Delta_{\mu\nu}^0(k) =
\left(\delta_{\mu\nu} - \frac{k_\mu k_\nu}{k^2}\right)\frac{1}{k^2}\,,
\label{eq:effective-gluon}
\end{equation}
where the flavour dependence of $\mathcal{I}_{\! ff'}(k^2)$ in \1eq{eq:I_ff}
enters through the quark dressing function $A_f(q^2)$.
With the kernel of Eq.~\eqref{eq:effective-gluon}, the dressed quark propagator
for each flavour $f=u,d,s$ is obtained from the gap equation
\begin{equation}
  S_{\!f}^{-1}(p) = Z_2\,(i\slashed{p} + m_f^R)
  + \underbrace{C_{\!\s F} \!\!\int_k\,
\gamma_{\alpha} S_{\!f}(k+p)\gamma_{\beta} 
\,D_{\! ff}^{\alpha\beta}(k)}_{\Sigma_f(p^2)} \,,
  \label{eq:gap-equation}
\end{equation}
where $C_F=4/3$ is the Casimir eigenvalue of the fundamental representation, 
and $\Sigma_f(p^2)$ denotes the quark self-energy. In addition, $m_f^R$ denotes the renormalized current mass, while $Z_2$ is the quark renormalization constant, which will be determined via the momentum subtraction renormalization scheme (MOM) \cite{Celmaster:1979km, Hasenfratz:1980kn, Braaten:1981dv, Athenodorou:2016oyh}. The inverse quark propagator is decomposed as
\begin{equation}
    S_{\! f}^{-1}(p) = i \slashed{p} A_f(p^2) + B_f(p^2) \,,
\end{equation}
and the renormalization-group invariant quark mass function is defined as $\mathcal{M}(p^2) = B(p^2)/A(p^2)$.

Furthermore, quark-antiquark bound states are described by the homogeneous BSE
\begin{equation}
{\cal A}_{ff'}(p_1,p_2) =
-\int_{k} \gamma_{\mu}\,
S_{\! f}(k_1)\,{\cal A}_{ff'}(k_1,k_2)\,S_{\! f'}(k_2)\,
\gamma_{\nu}\,D_{ff'}^{\mu\nu}(k)\,,
\label{eq:BSEI}
\end{equation}
with $k_i = k + p_i$ for $i=1,2$. It is convenient to introduce the total and relative momenta $P = p_1 - p_2$ and $p = (p_1 + p_2)/2$, in terms of which one writes ${\cal A}_{ff'}(p_1,p_2)\equiv{\cal A}_{ff'}(p,P)$. For a pseudoscalar meson, we employ the Dirac decomposition
\begin{equation}
{\cal A}_{ff'}(p,P) =
\left(
\chi_{1}^{ff'} + i\,\chi_{2}^{ff'}\!\slashed{P}
+ i\,\chi_{3}^{ff'}\!\slashed{p}\,(p\!\cdot\!P)
+ \chi_{4}^{ff'}\![\slashed{p},\slashed{P}]
\right)\gamma_{5}\,,
\label{eq:thechis}
\end{equation}
where the dressing functions $\chi_i^{ff'}(p^2,p\!\cdot\!P,P^2)$
encode the momentum dependence of the amplitude.
Finally, the coupling of the external baryonic current to a dressed quark is described by the $B-$vertex, $\Gamma_{\! B,f}^{\mu}(p_1,p_2)$, which satisfies an SDE of the form,
\begin{equation}
  \Gamma_{\! B,f}^\mu(p_1,p_2) =  \gamma^\mu
  - C_F\!\!\!\int_k \gamma_\alpha\, S_f(k_1)\, \Gamma_{\! B,f}^\mu(k_1,k_2)\, S_f(k_2)\,
  \gamma_\beta\, D_{\alpha\beta}^{ff}(k)\,.
  \label{eq:baryon-vertex-BSE}
\end{equation}

The $U(1)_B$ quark-number symmetry entails that the $B-$vertex satisfies the corresponding vector WTI,
\begin{equation}
  q_\mu\, \Gamma_{\! B,f}^\mu(p_1,p_2) =
  i\, S_f^{-1}(p_2) - i\, S_f^{-1}(p_1)\,.
  \label{eq:WTI-baryon}
\end{equation}
We emphasize that the 
present truncation guarantees the validity 
of the WTI in \1eq{eq:WTI-baryon}, essentially due to 
the fact that 
the defining equations for $S$ and 
$\Gamma_{B,f}^\mu $ (i and (iii) in \fig{fig:truncation}), share a common interaction kernel 
(blue band).

As was done in the electromagnetic case \cite{Miramontes:2025vzb}, the current-vertex can be decomposed as 
\begin{equation}
  \Gamma_{\! B,f}^\mu(p_1,p_2) = \Gamma_{\! B,f}^{\mu,\text{BC}}(p_1,p_2)
  + \Gamma_{\! B,f}^{\mu,\text{T}}(p_1,p_2)\,,
  \label{eq:GammaB-decomposition}
\end{equation}
where the so-called "Ball-Chiu" part,
$\Gamma_{\! B,f}^{\mu,\text{BC}}$ saturates the 
WTI of \1eq{eq:WTI-baryon} \cite{Ball:1980ay}, while the remainder, 
$\Gamma_{\! B,f}^{\mu,\text{T}}$, 
vanishes when contracted by $q_\mu$. 
Then, $\Gamma_{\! B,f}^{\mu,\text{BC}}$ may be 
expressed solely in terms of the quark dressing functions $A_f$ and $B_f$,  while the transverse components encode dynamical information beyond the WTI constraints. 
In particular, $\Gamma_{\! B,f}^{\mu,\text{T}}$ develops the vector-meson poles generated by the quark–antiquark scattering kernel, namely, the $\rho$ in isovector channels and the $\omega$ in the isoscalar \cite{Miramontes:2019mco, Miramontes:2021xgn, Williams:2018adr}; these contributions are crucial 
for the behaviour of the form factor and its charge radius.
Within the present SDE-BSE framework, these poles emerge dynamically from the same scattering kernel that describes the bound-state spectrum, rather than being introduced through a vector-meson-dominance Ansatz \cite{OConnell:1995nse}.
For the pion BFF, this becomes relevant once isospin breaking ($m_u \neq m_d$) is included. 
Indeed, at that point,  
the matrix element becomes sensitive to the isoscalar vector strength encoded in $\Gamma_{\! B,f}^{\mu,\text{T}}$, most notably that associated with the $\omega$ resonance.

The tensor basis of the vertex $\Gamma_{B,f}^\mu$
employed in this work
is identical to the one used  in \cite{Miramontes:2025vzb}
for the quark-photon vertex [see Eqs. (4.4) and (4.7) therein].

\section{Results}
\label{sec:results}
\begin{figure*}[t!]
\centerline{%
\includegraphics[width=1.05\textwidth]{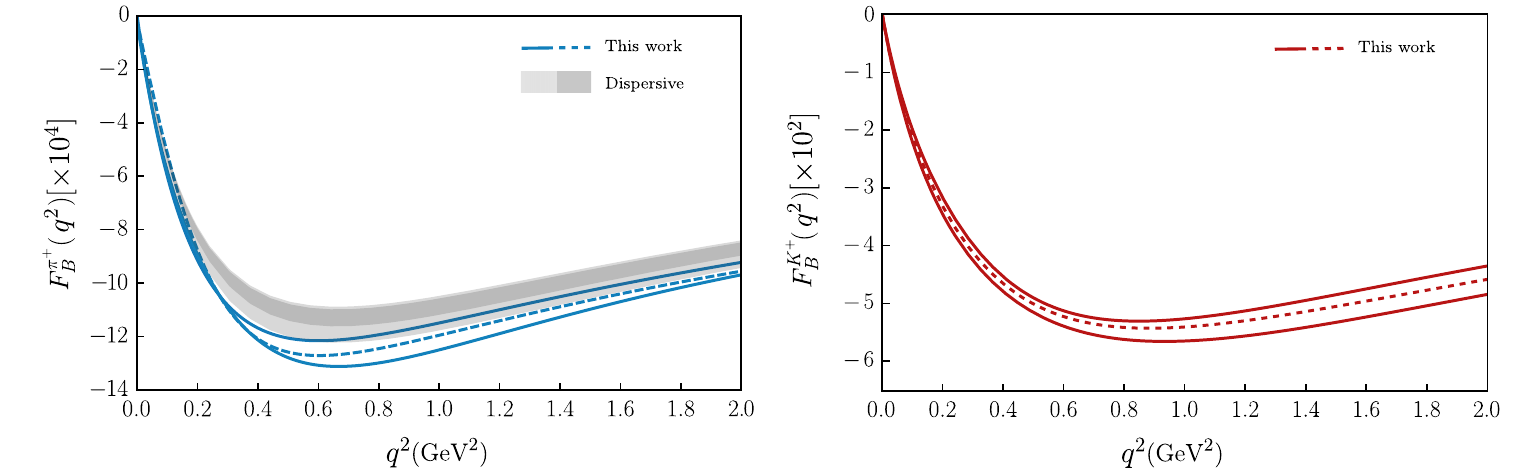}}
\caption{Left panel: Computed baryonic form factor for the charged pion, the dashed line corresponds to the central parametrization of the $\tilde{\alpha}_T$ while the solid lines are the variations. The results are compared to the dispersive extraction from the BaBar and KLOE data in \cite{Sanchez-Puertas:2021eqj}. Right panel: Corresponding results for the charged kaon.}
\label{fig:BFF}     
\end{figure*}
In this section, we present the numerical results for the computed quark propagators, meson masses and baryonic form factors. For a detailed description of the numerical techniques we refer the reader to \cite{Sanchis-Alepuz:2017jjd}.\\

(${\it i}$) \textbf{Numerical inputs.} The quark gap equations are renormalized in the MOM scheme at  $\mu = 4.3~\text{GeV}$. The current quark masses are fixed to
\begin{equation}
  m_u = 3.7~\text{MeV}\,, 
  \quad
  m_d = 6.2~\text{MeV}\,,
  \label{eq:input1}
\end{equation}
which reproduce the physical charged pion mass and its decay constant. 
In addition, the strange-quark mass is chosen as
\begin{equation}
  m_s = 95~\text{MeV}\,,
  \label{eq:input2}
\end{equation}
such that the charged kaon mass is correctly described within the same flavour-dependent interaction framework. The effective interaction $\mathcal{I}_{ff'}(k^2)$ and its parametrization are taken from Ref.~\cite{Gao:2024gdj} without further adjustment. Thus, no additional parameters are introduced in the present study beyond the standard light-quark masses.
\\

(${\it ii}$) \textbf{Quark propagators.} For each flavour $f = u,d,s$ we first solve the corresponding quark gap equation, obtaining the dressing functions $A_f(p^2)$, $B_f(p^2)$ and the mass function $\mathcal{M}_f(p^2)$. In Euclidean space-time, the total momentum of an on-shell bound state of mass $M_{\textbf{s}}$ can be chosen as
\begin{equation}
  P = (0,0,0,iM_{\textbf{s}})\,,
  \qquad P^2=-M_{\textbf{s}}^2\,.
\end{equation}
With this parametrization, the squared momenta flowing through the quark and
antiquark lines are
\begin{equation}
  k_{1,2}^2 = k^2 + \frac{P^2}{4} \pm k\!\cdot\! P\,.
\end{equation}
As the integration momentum $k$ is varied, the quantities $k_{1,2}^2$
describe parabolic domains in the complex $k_{1,2}^2$-plane, which implies that the
quark dressing functions entering $S(k_{1,2})$ must be known in the complex region.

The required analytic continuation of the quark propagator to the complex plane is implemented employing the Cauchy interpolation method described in \cite{Sanchis-Alepuz:2017jjd,Fischer:2005en,Krassnigg:2008bob}.

The resulting mass functions $\mathcal{M}(p^2)=B(p^2)/A(p^2)$ are displayed in the right panel of \Cref{fig:aT_mass} for $f=u,d,s$. In all cases, $\mathcal{M}(p^2)$ shows the expected infrared enhancement from dynamical chiral symmetry breaking, and evolves smoothly towards the current masses at large momentum. The corresponding central values at the origin are $\mathcal{M}_u(0)= 0.328$ GeV, $\mathcal{M}_d(0)= 0.341$ GeV and $\mathcal{M}_s(0)= 0.591$ GeV. The band around the central lines corresponds to the error propagation from the variation of the $\tilde{\alpha}_T(k^2)$, displayed in the left panel of \Cref{fig:aT_mass}.
\\

(${\it iii}$) \textbf{Masses and decay constants.} The quark propagators obtained from the gap equations are then inserted into the
homogeneous BSEs for the pion and kaon. For a given value of the total momentum
$P^2$, the BSE defines an integral equation for the BS amplitude, which can be expressed as an eigenvalue problem. In this formulation, one determines an eigenvalue $\lambda(P^2)$ associated with the BSE kernel, and physical bound states are identified at those points $P^2=-M^2$ where the leading eigenvalue fulfills $\lambda(P^2)=1$. The computed BS amplitudes are subsequently employed in the impulse-approximation evaluation of the baryonic current, and the extraction of the BFF.

Using the inputs for the quark masses in \1eq{eq:input1} and \1eq{eq:input2}, we obtain the following mass and decay constant for the pion,
\begin{equation}
    m_{\pi} = 0.139 \text{GeV}, \qquad  f_{\pi} = 0.130 \text{GeV}
\end{equation}
while for the kaons we have,
\begin{align}
     & m_{K^{\pm}} = 0.493 \text{GeV}, \qquad  f_{K^{\pm}} = 0.157 \text{GeV} \\
     & m_{K^0} = 0.499 \text{GeV}, \qquad  f_{K^0} = 0.157 \text{GeV}
\end{align}
In comparison, the corresponding experimental values are $m_{\pi} = 0.139$GeV and $f_{\pi} = 0.130$GeV for the pion, and $m_K^{\pm} = 0.493$GeV, $f_K^{\pm} = 0.155$~GeV, $m_K^{0} = 0.497$GeV, $f_K^{0} = 0.155$GeV for the kaons \cite{ParticleDataGroup:2024cfk}.
The choice of current-quark masses is guided by the requirement that, within the present approach, the charged pseudoscalar sector be reproduced. Within this approach, the resulting $K^\pm$-$K^0$ splitting reflects explicitly the strong isospin-breaking effect driven by $m_d-m_u$. The small deviation of $m_{K^0}$ from its physical value, of around $0.5\%$, is not unexpected, because the physical $K^\pm$-$K^0$ mass difference receives a repulsive 
electromagnetic contribution to $m_{K^\pm}$ that is absent here \cite{FlavourLatticeAveragingGroupFLAG:2024oxs}.
It is expected that 
the inclusion of electromagnetic isospin breaking at the level of the SDE-BSE system would shift the neutral-kaon mass toward its physical value,  
yielding the full splitting \cite{Miramontes:2022mex}.
\\

(${\it iv}$)  \textbf{Baryonic form factor.}
With the quark propagators and meson BS amplitudes computed, the baryonic form
factor can be obtained from the impulse-approximation expression of \1eq{eq:JB-impulse}. The remaining ingredient in this current matrix element is the dressed $B-$vertex,
$\Gamma_{\! B,f}^\mu$, which we determine by solving the SDE 
of \1eq{eq:baryon-vertex-BSE}
in the space-like region, for each flavour $f=u,d,s$. We compute the full vertex in its most general form, retaining all 12 dressing functions, and solve the vertex equation for the set of $q^2$ values at which the form factor is evaluated.  After projection onto the 12-dimensional Dirac basis and discretization of the momentum variables, the SDE reduces to a linear system for the vertex dressing functions. This system is solved by standard matrix-inversion techniques (LU decomposition \cite{Press:1992zz}); the form factors obtained coincide with those 
shown in Figs. 6 and 7 of 
\cite{Miramontes:2025vzb}, and will not be repeated here. 
The resulting vertex, together with the propagators and BS amplitudes, is then inserted into \1eq{eq:JB-impulse} to compute the baryonic current and extract $F^{\textbf{s}}_{\! B}(q^2)$ from \1eq{eq:FB}. 
\begin{figure}[t!]
\centerline{%
\includegraphics[width=0.5\textwidth]{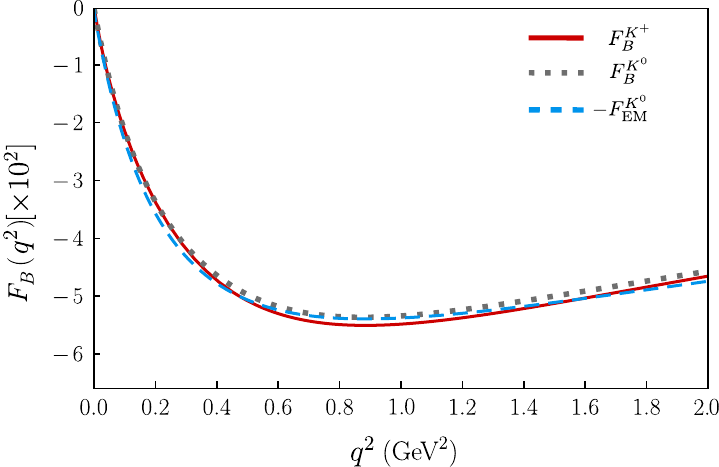}}
\caption{Baryonic form factor for the neutral kaon (dotted line) compared with the charged one (solid line) and to the corresponding neutral-kaon electromagnetic form factor from \cite{Miramontes:2025vzb} (dashed line).}
\label{fig:K0}     
\end{figure}

For the BFF of the charged pion, a useful benchmark is provided by the data-driven analysis of \cite{Sanchez-Puertas:2021eqj}, where the form factor was extracted indirectly from high-statistics $e^+e^-\to\pi^+\pi^-$ data from BaBar and KLOE. Since a direct experimental probe of the baryon current is not available, this procedure yields a phenomenologically constrained determination of $F^{\pi^+}_{\! B}(q^2)$ and its slope at the origin.  

Turning now to our main results, the left panel of \Cref{fig:BFF} shows the BFF of the charged pion, $F^{\pi^+}_B(q^2)$, in the space-like region. The form factor vanishes at the origin, and remains numerically small over the entire momentum range shown. In the same figure we compare our calculation to the dispersive extraction of \cite{Sanchez-Puertas:2021eqj}, obtained from the BaBar \cite{BaBar:2009wpw} and KLOE \cite{KLOE:2004lnj, KLOE:2008fmq, KLOE-2:2017fda, KLOE:2010qei} data and analytically continued to the space-like regime; 
our results 
correspond to the three blue curves. The central curve
corresponds to the reference parametrization of the modified Taylor
coupling $\tilde{\alpha}_T(k^2)$, see 
left panel of \fig{fig:aT_mass},  
while the upper/lower curves are obtained by varying $\tilde{\alpha}_T
(k^2)$ within the red band shown in 
the same figure. 
The difference between our results and those of the  dispersive analysis may be attributed 
to truncation effects, as well as 
isospin-breaking contributions 
not captured by the present approach, most notably electromagnetic effects.

From the slope of $F_{\!B}^{\pi^+}(q^2)$ at the origin, we extract the baryonic radius. Using the central interaction and the two variations of the
modified Taylor coupling, we obtain
\begin{equation}
  \langle r_{\! B}^2\rangle^{1/2}_{\pi^+}
  = 0.043(2)\ {\rm fm}\,,
\end{equation}
which is in very good agreement with the data-driven estimate of 
\mbox{$\langle r_{\!B}^2\rangle^{1/2}_{\pi^+} = 0.041(1)$ fm}.

The charged kaon BFF is shown in the right panel of \Cref{fig:BFF}. It satisfies $F_B^{K^+}(0) = 0$, and its magnitude for $q^2>0$ is significantly enhanced with respect to the pion case, reflecting the increased flavour asymmetry in the $u\bar s$ system \cite{Broniowski:2021awb}. As before, the three curves correspond to the variation of the $\tilde{\alpha}_T(k^2)$. The resulting spread is moderate over the full $q^2$ range, indicating that the qualitative behaviour is stable within the present interaction uncertainty. 
\begin{table}[t!]
\centering
\renewcommand{\arraystretch}{1.15}
\begin{tabular}{|l|c|c|c|}
\hline
 Radius (fm) & $\pi$ & $K^{\pm}$ & $K^{0}$ \\
\hline
This work     & 0.043(2)  & 0.265(7) & 0.262(7) \\
\hline
BaBar  \cite{Sanchez-Puertas:2021eqj}       & 0.041(1)  & --      & --      \\
\hline
KLOE  \cite{Sanchez-Puertas:2021eqj}        & 0.041(1)  & --      & --      \\
\hline
Yukawa model \cite{Sanchez-Puertas:2021eqj} & 0.040     & --      & --      \\
\hline
Chiral NJL   \cite{Broniowski:2021awb}        & 0.06(1)   & 0.24(1)  & 0.23(1)  \\
\hline
\end{tabular}
\caption{Summary of baryonic radii $\langle r_B^2 \rangle^{1/2}$ (in fm).}
\label{tab:radii}
\end{table}
Using the low-$q^2$ behaviour, we determine the charged-kaon baryonic radius,
\begin{equation}
  \langle r_{\!B}^2\rangle^{1/2}_{K^+}= 0.265(7)\ {\rm fm}\,,
\end{equation}
where the uncertainty quantifies the sensitivity to the effective interaction
used in our error analysis. The comparatively larger value of
$\langle r_{\! B}^2\rangle^{1/2}_{K^+}$ indicates that the kaon channel enhances the
baryonic-structure signal.

In \Cref{fig:K0} we compare the 
BBFs of the charged
and neutral kaons. A particularly noteworthy feature is that the electromagnetic form factor, $-F_{\rm EM}^{K^0}(q^2)$, closely follows the baryonic kaon form factors. A way to understand this is to decompose both observables into
flavour contributions with different external-current weights. For the neutral
kaon, $K^0=d\bar s$, the electromagnetic current couples with
$Q_d=-1/3$ and $Q_{\bar s}=+1/3$, so that, schematically,
\begin{equation}
F_{\rm EM}^{K^0}(q^2) = Q_d\,F_d(q^2)+Q_{\bar s}\,F_{\bar s}(q^2)
= \frac{1}{3}\Big[F_{\bar s}(q^2)-F_d(q^2)\Big]\,,
\end{equation}
and therefore
\begin{equation}
-\,F_{\rm EM}^{K^0}(q^2) = \frac{1}{3}\Big[F_d(q^2)-F_{\bar s}(q^2)\Big]\,.
\end{equation}
In contrast,  in the baryon-number current, $j_B^\mu=\frac{1}{3}\sum_f \bar q_f\gamma^\mu q_f$,  
the quark ($d$) and antiquark ($\bar s$) enter 
with opposite baryon number, $B_d=+1/3$ and $B_{\bar s}=-1/3$. Consequently one may write, in the same schematic notation,
\begin{equation}
F_{B}^{K^0}(q^2)=B_d\,F_d(q^2)+B_{\bar s}\,F_{\bar s}(q^2)
=\frac{1}{3}\Big[F_d(q^2)-F_{\bar s}(q^2)\Big]\,,
\end{equation}
which makes the relation $F_B^{K^0}(q^2) =  -F_{\rm EM}^{K^0}(q^2)$ manifest.

For this case, the neutral-kaon baryonic radius,
\begin{equation}
  \langle r_{\! B}^2\rangle^{1/2}_{K^0} = 0.262(7)\ {\rm fm}\,,
\end{equation}
is strikingly close to the charged case, and compares well with the electromagnetic radius of the neutral kaon
obtained in \cite{Miramontes:2025vzb}, $\langle r_{\rm EM}^2\rangle^{1/2}_{K^0}=0.270\ {\rm fm}$.
We remark, however, that the small residual difference between the neutral-kaon electromagnetic and baryonic results is primarily of a technical origin: the electromagnetic form factor in \cite{Miramontes:2025vzb} was computed in the
isospin-symmetric limit, whereas the present BFF calculation is performed in an explicitly isospin-breaking setup. Consequently, the $d$-quark mass (and hence the dressed propagator), as well as the neutral kaon BS amplitude entering the electromagnetic calculations, are not identical to those employed here, which leads to a slight mismatch in the extracted radii and curves. Finally, we summarize our radii results in \Cref{tab:radii}. For the kaons, where no dispersive benchmark is currently available, our results are of the same qualitative size as those found in chiral quark model studies \cite{Broniowski:2021awb}, and support the picture that the baryonic distribution in the kaon is substantially more extended than in the pion.

\section{Conclusions} \label{sec:outlook}

We have computed the baryonic form factors of $\pi^+$, $K^+$ and $K^0$ in an explicitly strong isospin-breaking scenario, and extracted the corresponding radii from the low-$q^2$ behaviour.
Our central results are
$\langle r_B^2 \rangle_{\pi^+} = (0.043(2)\,\mathrm{fm})^2$,
\mbox{$\langle r_{\! B}^2 \rangle_{K^{+}} = (0.265(7)\,\mathrm{fm})^2$}, and
$\langle r_{\! B}^2 \rangle_{K^{0}} = (0.262(7)\,\mathrm{fm})^2$.
The pion value is consistent with the available dispersive benchmark, while the kaon radii are significantly larger and compare with the corresponding charge radius of the electromagnetic counterpart. 
The low-momentum behaviour of these form factors is controlled by the interplay between 
symmetry-imposed constraints and the dynamical vector-meson content encoded in the "B-vertex", 
which is generated from its own SDE.
Let us finally mention that electromagnetic corrections to the baryonic form factors and their radii
may be obtained from 
the present approach 
by adding photonic loops in the 
dynamical equations of 
\fig{fig:truncation}, and carrying out 
an appropriate re-tuning 
of $m_u$, $m_d$ and $m_s$, see e.g., \cite{Miramontes:2022mex}.

\section*{Acknowledgements}
The authors thank P. Sanchez-Puertas, E. Ruiz-Arriola, and W. Broniowski  for useful communications. 
This work is funded by the Spanish MICINN grants PID2020-113334GB-I00 and
PID2023-151418NB-I00,  
the Generalitat Valenciana grant CIPROM/2022/66,
and CEX2023-001292-S by MCIU/AEI.

\bibliographystyle{elsarticle-num-names}
\bibliography{bibliography.bib}

\end{document}